\documentclass[11pt,twoside]{article}  
\usepackage{asp2006}
\usepackage{adassconf}
\usepackage{graphicx}

\begin{document}   

%
%

\paperID{O13.2}

%

\title{A VO-driven Astronomical Data Grid in China }

%
%
%
%
%

\markboth{CUI et al.}{VO-driven Astronomical Data Grid}

%
%
%
%

\author{Chenzhou CUI, Boliang HE, Yang YANG, Yongheng ZHAO}
\affil{National Astronomical Observatories, CAS}

%

\contact{Chenzhou CUI}
\email{ccz@nao.cas.cn}

%
%
%

\paindex{CUI, C.}
\aindex{HE, B.}
\aindex{YANG, Y.}
\aindex{ZHAO, Y.}

%

\keywords{archives!collection management, archives!services, catalogs, data!grids, data management}


\begin{abstract}          
With the implementation of many ambitious observation projects, including LAMOST, FAST, and Antarctic observatory at Doom A, observational astronomy in China is stepping into a brand new era with emerging data avalanche. In the era of e-Science, both these cutting-edge projects and traditional astronomy research need much more powerful data management, sharing and interoperability. Based on data-grid concept, taking advantages of the IVOA interoperability technologies, China-VO is developing a VO-driven astronomical data grid environment to enable multi-wavelength science and large database science. In the paper, latest progress and data flow of the LAMOST, architecture of the data grid, and its supports to the VO are discussed.
\end{abstract}

%
%

\section{Introduction}
Nowadays, astronomers are facing data avalanche sourced from projects like sky surveys, time-domain observations, large scale numerical simulations, and so on, which are powered by large scale data acquisition equipments and high performance computing abilities. China is on the track to fast change. Chinese astronomers are just in the eve of data flood. With the implementation of many ambitious observation projects, including Large Sky Area Multi-Object Fiber Spectroscopic Telescope (\htmladdnormallinkfoot{LAMOST}{http://www.lamost.org/}), Five hundred meter Aperture Spherical radio Telescope (FAST), Antarctic observatory at Doom A, et al, tens of terabyte scientific data will be there waiting for the community to mine.
Virtual Observatory (VO) is a data-intensively online astronomical research and education environment, taking advantages of advanced information technologies to achieve seamless, global access to astronomical information. VO aims to make multi-wavelength science and large database science as seamless as possible, and it will act as a facility class data infrastructure (Lawrence 2009). The organization to coordinate global development on the VO and cultivate its awareness and usage is the International Virtual Observatory Alliance (\htmladdnormallinkfoot{IVOA}{http://www.ivoa.net/}).
Chinese Virtual Observatory (China-VO) is the national VO project in China initiated in 2002 (Cui \& Zhao 2004). To provide such a facility class data infrastructure for Chinese astronomical archives, the China-VO is focusing on developing a VO-driven astronomical data grid. In the following sections, we first give out the data flow of the LAMOST, then describe the architecture of the data grid. Its supports to the VO will be discussed in the last section.
\section{LAMOST Data Flow}
LAMOST is an innovative reflecting Schmidt telescope with its active reflecting corrector of 5.72m x 4.40m and primary mirror of 6.67m x 6.05m. The overall concept and key technical innovations makes it a unique astronomical instrument in combining a large aperture with a wide field of view. The available large focal plane accommodates up to 4000 fibers, by which the collected light of distant and faint celestial objects down to 20.5 magnitude is fed into the spectrographs, promising a very high spectrum acquiring rate of several ten-thousands of spectra per night. LAMOST is located at the XingLong Station of National Astronomical Observatories of China, as a national facility open to the whole community. In Aug. 2008, the installation and assembly of all telescope hardware systems were completed. Calibration observation started in 2009.
\begin{figure}
  \includegraphics[width=60mm]{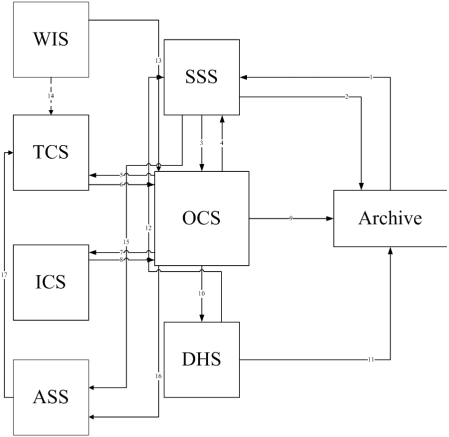}\\
  \caption{Data flow of the LAMOST}\label{O13.2-fig-1}
\end{figure}
Fig. 1 shows the data flow of the LAMOST. Data and information derived from seven software systems, including Sky Survey Strategy system (SSS), Observation Control System (OCS), Data Handling System (DHS), Environments Information System (EIS), Telescope Control System (TCS), Instruments Control System (ICS), and Astrometry Support System (ASS), are fed into Archive system finally. At a typical observation night, about 30 gigabyte data will be collected by the telescope through its 16 spectrographies and 32 CCD cameras. By the end of a 5-year sky survey, final data size of the archive is about 30 to 50 TB.
\section{Data Grid Architecture}
The China-VO data grid is aiming to act as a facility class data infrastructure for these archives mentioned above, providing easy data management functions for system administrators and standard data access services for scientific users.
\subsection{Logical Architecture}
\begin{figure}
  \includegraphics[width=80mm]{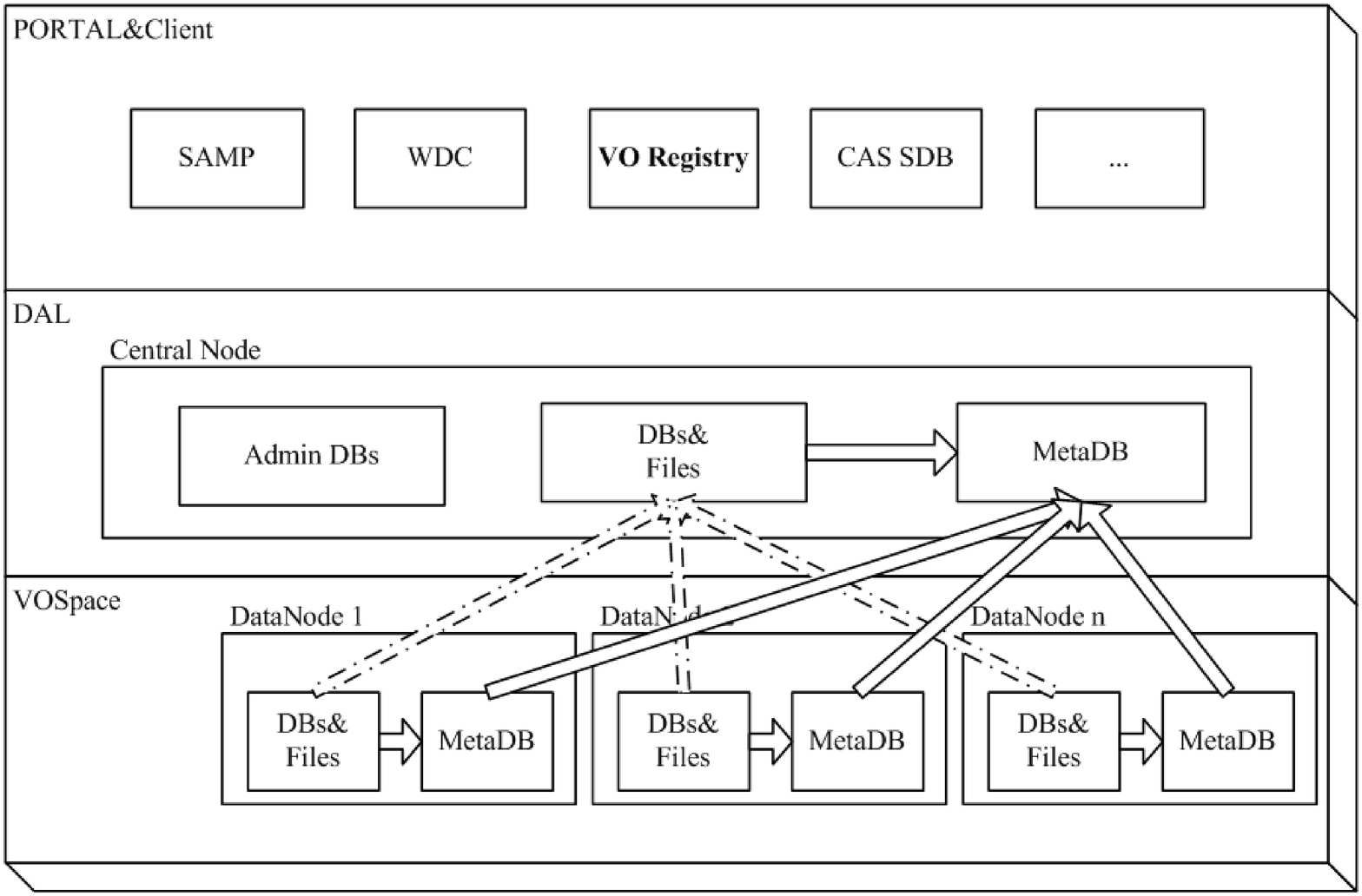}\\
  \caption{Logical architecture of the data grid}\label{O13.2-fig-2}
\end{figure}
It is a distributed data grid, with a logical architecture shown in fig. 2. The system consists of 3 layers, VOSpace specification complimented storage layer, IVOA DAL complimented data access layer, and various of portals and clients. Datasets are composed of two parts, data and metadata. Data is stored in one of two ways, database or file. Metadata is managed by relational database systems. In the data grid, there is a central metadata database (MetaDB), which is the first object queried for data discovery and data access. Metadata information locating at different DataNodes is harvested periodically to the central MetaDB. Data body of these datasets are stored in different data centers and observatories.
\subsection{Technical Architecture}
Fig. 3 shows the technical architecture. Combining \htmladdnormallinkfoot{iRODS}{https://www.irods.org/} with VOSpace, a VO-oriented data grid is designed. It will be used for storage resource virtualization, data management, synchronization, backup and recover, and file level data access. Metadata and catalogs are stored in open source database systems, i.e. MySQL and PostgreSQL. A RESTful Data Access Layer (DAL) will convert IVOA standard data access queries into native queries for low-lever database systems and return responses in IVOA standard formats. Popular IVOA DAL interfaces, for example, TAP, SSAP, SIAP and simple cone search, are supported. In order to support large scale of data quires, asynchronous service is designed. Rather than using sessions, query states are maintained in database tables. For data queries requiring two and more steps, including SSAP and SIAP, MetaDB and catalogs will be queried first through TAP interfaces and then retrieve the actual results as data files through VOSpace and iRODS interfaces.
\section{Supports to the VO}
It is a VO-driven and VO-oriented data grid. Many IVOA specifications and related technologies are adopted. Its supports to the VO can be summarized in three categories.
First, on data encoding and metadata format, query responses are encoded into VOTable files; metadata is following the IVOA Resource Metadata (RM) schema. Second, IVOA standard data access interfaces are supported, including TAP and Cone Search for catalogs, SSAP for spectrum, SIAP for images, and VOSpace for file access. Third, IVOA Registry and SAMP are supported to expose archives in the data grid to the whole VO community and interlink various of applications.
\acknowledgments
The China-VO project is supported by NSFC (60603057, 10778623, 10820002, 60920010), MOST (China) (2006AA01A120), BMS\&TC (2007A085) and CAS (INFO-115-C01-SDB3-04).
\begin{figure}
  \includegraphics[width=70mm]{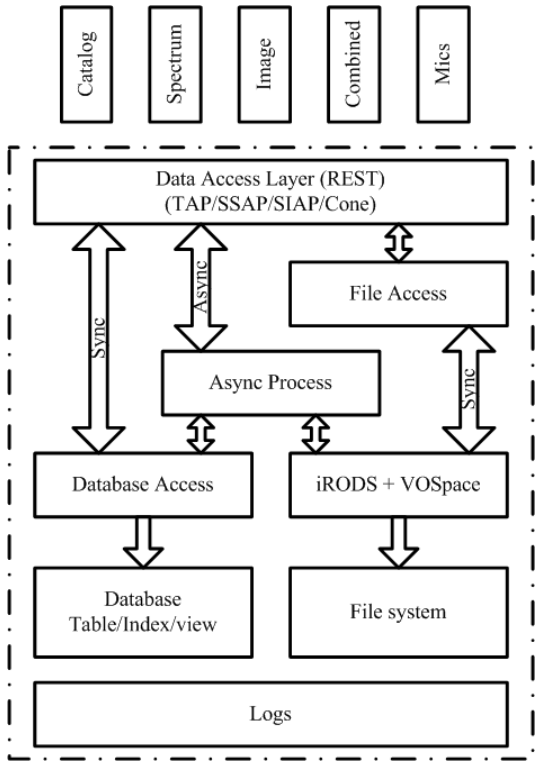}\\
  \caption{Technical architecture of the data grid}\label{O13.2-fig-3}
\end{figure}

\end{document}